\documentclass[10pt,nobibnotes,preprint,injp,nofootinbib,showpacs]{article}
\usepackage{latexsym}
\usepackage{amssymb}
\usepackage{amsmath}
\usepackage{amscd}
\usepackage{amsthm}
\usepackage{times}
\usepackage{mathrsfs}
\usepackage[left=2cm,top=2.5cm,right=2.5cm,bottom=1.5cm]{geometry}
\usepackage{amsfonts}
\usepackage{braket}
\usepackage{color}
\usepackage{graphicx}
\usepackage{bm}
\usepackage{textcomp}
\usepackage{setspace}
\usepackage{bm}

\makeatletter
\def\@xfootnote[#1]{%
  \protected@xdef\@thefnmark{#1}%
  \@footnotemark\@footnotetext}
\makeatother
\begin{document}
\begin{spacing}{2}
\def\Zz{\mathbb{Z}}
\def\T{\hat{T}}
\begin{center}
\large{\bf{Effect of Temperature on the Complexity of Solid Argon System}} \\
\vspace{3mm}
\normalsize{ A Giri$^{1}$\footnote[*]{Corresponding author, E-mail : amal.vu@gmail.com}, S Dey$^1$ and P Barat$^{1}$ } \\
\vspace{3mm}
\normalsize{$^1$ Variable Energy Cyclotron Centre, 1/AF Bidhannagar, Kolkata, West Bangal, India, 700064}\\
\end{center}
\vspace{1cm}
{\bf Abstract:} \hspace{2mm}We study the measure of complexity in solid Argon system from the time series data of kinetic energy of single Argon atoms at different equilibrated temperatures. To account the inherent multi-scale dependence of the complexity, the multi-scale entropy of the time series of kinetic energy of individual Argon atoms are computed at different equilibrated temperatures. The multi-scale entropy study reveals that the dynamics of an atom becomes more complex at higher temperatures and the result corroborates well with the variation of the pair correlation function of the atoms in the solid Argon crystal. Also, we repeat the multi-scale entropy analysis for program generated Levy noise time series and for time series data obtained from the outcomes of exponential decay with noise $\,d x(t)=-x(t)\,dt+\sigma \,dB(t)$ (Langevin equation). Our study establishes that the scale dependence of sample entropy for time series of kinetic energy of individual atoms in solid Argon system has similar tendency as that of Levy noise time series and the outcomes of exponential decay with noise (Langevin equation).\\\\
{\bf Keywords:} \hspace{2mm}Complexity and complex systems; Entropy and other measures of information; Equilibrium thermodynamics\\\\
{\bf PACS Nos:}\hspace{2mm} 05.45.Tp; 05.40.-a; 05.40.Ca\\
\def\be{\begin{equation}}
\def\ee{\end{equation}}
\def\bea{\begin{eqnarray}}
\def\eea{\end{eqnarray}}
\def\zbf#1{{\bf {#1}}}
\def\bfm#1{\mbox{\boldmath $#1$}}
\def\hf{\frac{1}{2}}
\smallskip
\newpage 
\noindent
{\bf 1.  Introduction}\\
Most of the systems observed in nature are complex in nature. Examples of complex systems are the flow of water in a river, metabolic activities in  our body, the dynamic behavior of body cells etc. The manifestation of the dynamics of these complex systems are in the time and spatial scale. They are complex to us as because we can not predict their behavior or model them by a proper mathematical formalism. Modern science is aiming to understand macroscopically the complex behavior of these systems by various statistical tools. The time series data, associated to an output variable, generated by such complex systems generally contain deterministic and stochastic components. Stochastic component represents the fluctuation in the output variable. The presence of this fluctuation is not simply due to contaminated noise. Rather a signature of the underlying dynamics of the system is reflected in the fluctuation of such uncontaminated stochastic component. Generally for time series analysis two classical approaches are used and these are related to deterministic and stochastic mechanism \cite{Takens}. Both of them can explain the underlying dynamics of the system in a complementary manner. Instead of using any particular mechanism we emphasize on a method which can quantify the degree of complexity of the time series. The measured complexity can be used to discriminate different time series generated by different dynamical systems or by same dynamical system having different physical conditions. In this context it is most important to mention the work of Costa \emph{et~al.} \cite{Costa1} that is the first of its kind to define a dynamical system in terms of its complexity. Subsequently several works have been reported mostly on biological systems [3-10] to measure the complexity of the dynamical systems.

The crystalline system is a very ordered structure as revealed by X-ray diffraction technique. In equilibrium at a fixed temperature the velocity of its constituents obey a stable Boltzmann distribution. Apparently one may think that such a system is one of the simplest one. However, if we go in deep and try to understand the dynamics of the system in terms of its constituent atoms, i.e. if we reduce the length scale and try to observe  the dynamics of individual constituents atoms it will obviously be complex. The atoms in crystalline systems can be of various types forming a cluster around a lattice point in the form of basis atoms. Interaction among these atoms is nevertheless highly non-linear. The atoms in a crystal at equilibrium are not static. They vibrate about their mean positions and the vibration frequencies  are different for different atoms even though they follow a distribution of frequency. In addition the kinetic energy (KE) possessed by an atom of the system can have different values with finite probability. In spite of the fact that the system is in thermodynamic equilibrium, the atoms at every moment acquire fresh KE value. The atoms vibrate about their mean position since they possess KE but don't have adequate space to move on in the solid structure although and also their vibration are not independent \cite{Debye}. Thus, these apparent-simple systems are not as simple as it comes in ones thought. Moreover, for higher equilibrium temperatures the constituent atoms vibrate with higher amplitude about their mean making the atom dynamics much more complicated. In crystalline system the interaction with nearer neighbors leads to deterministic component and stochastic component of the force field turns up from the interaction with the distant atoms. In this work we consider solid Argon (Ar) systems at different equilibrium temperatures and with the help of molecular dynamics \cite{MD++} simulation the temperature dependence and multiple scale factor dependence of the complexity of such simplest system are discussed.\\\\
{\bf 2. Traditional entropy and complexity}\\
For every system we need certain amount of information to describe it. In case of complex system this information is represented by the quantity complexity. By calculating complexity from the time series of a dynamical variable of physical system we can conclude about the dynamics of the physical process. As per as information of a system is concerned, physicists are accustomed with traditional entropy representation of the system which provides the randomness or disorderedness of the system. It merely evaluates the appearance of repetitive patterns of a time series and also no straightforward relationship exists between the repetitive pattern of a time series and its complexity. Complexity is related to``meaningful structural richness" \cite{Pincus1} of the time series whereas the entropy based measurement looks for the randomness or the absence of regularity in a time series. Thus, for uncorrelated random noise entropy based method generates the highest value although the time series is not complex. Neither a completely predictable  nor a completely unpredictable signal is structurally rich and both of them are not complex. In contrast, the assigned entropy to predictable signal is minimum and it monotonically increases with randomness of the signal to reach the maximum value for uncorrelated random signal (white noise). Thus, in time series analysis, entropy calculation doesn't lead to the proper understanding about the complex nature of the time series. For instance, entropy based methods assign higher entropy values to certain pathologic biological processes that generate irregular outputs than to healthy biological that are acutely regulated by multiple interacting control processes \cite{Costa1,Costa2} although the loss of complexity is a generic feature of pathologic dynamics and the biological complexity monotonically degrades with aging and disease. This contrast indicates the need for a thematically faithful formalism, instead of traditional entropy based measurements, for general applications so that visual intuition matches numerical results, for broad classes of stochastic processes as well as for dynamical systems.\\\\
{\bf 3. Techniques of complexity measurement and introduction of multi-scale factor}\\
Structural richness of a time series is associated to the inherent multiple spatio-temporal scale of the complex system. Generally one tries to distinguish a chaotic complex system via parameter estimation. The parameters typically associated with chaotic complex systems are the measures of dimension, rate of information generated (entropy) and the Lyapunov spectrum. The classification of dynamical systems via entropy and the Lyapunov spectra stems from the works of Kolmogorov \cite{Kolmogorov}, Sinai \cite{Sinai} and Oseledets \cite{Oseledets}, though these works rely on ergodic theorems and the results are applicable to probabilistic settings. Dimension formulae are motivated by a construction in the entropy calculation and generally resemble Hausdorff dimension calculations. The mentioned theoretical works above are not intended as a means to effectively and appropriately discriminate dynamical systems, given the data is finite and noisy. Pincus \cite{Pincus1, Pincus9} has came with a solution by proposing a family of system parameters called approximate entropy (ApEn). It can potentially distinguish low-dimensional deterministic systems, periodic and multiply periodic systems, high-dimensional chaotic systems, stochastic, and mixed systems. \\\\
\textbf{\textit{Construction of approximate entropy}}: 

For any finite time series $\{\xi_i\}=[\xi_1,\xi_2,..,\xi_i,...\xi_N]$ of N data points a vector sequences $u(1)$ through $u(N-m+1)$, defined by $u(j)=[\xi_j,\xi_{j+1},..,\xi_{j+m-1}]$ with $1\leq j\leq (N-m+1)$ can be constructed. These vectors represent $m$ consecutive $\xi$ values, commencing with the \textit{j}th point. Define the distance $d[u(j),u(k)]$ between vectors $u(j)$ and $u(k)$ as the maximum difference in their respective scalar components. The vector sequence u(1),u(2), ... ,u(N-m+ 1) can be used to construct, for each  $j\leq (N-m+1)$, $P_j^{\prime m}(r)=$ (number of $j\leq (N-m+1)$ such that $d[u(j),u(k)]\leq r)/(N-m+1)$. The $P_j^{\prime m}(r)$'s measure within a tolerance $r$ the regularity, or frequency, of patterns similar to a given pattern of window length $m$. Define $U^m(r)=(N-m+1)^{-1}\sum_{j=1}^{N-m+1}\ln P_j^{\prime m}(r)$, where $\ln$ is the natural logarithm, then define the parameter $ApEn(m,r,N)=U^m(r)-U^{m+1}(r)$.

Mathematically, ApEn is the part of a general development as the rate of entropy for an approximating Markov chain to a process \cite{Pincus2}. In applications to heart rate, findings have discriminated groups of subjects via ApEn, in instances where classical [mean, standard deviation (SD)] statistics does not show clear group distinctions [19-23]. In applications to endocrine hormone secretion data based on as few as $N =72$ points, ApEn has provided vivid distinctions between actively diseased subjects and normals, with nearly 100\% specificity and sensitivity \cite{Pincus8}.

Informally, for $N$ points, the family of parameters $ApEn(m, r, N)$ is approximately equal to the negative average natural logarithm of the conditional probability that two sequences that are similar for $m$ points remain similar, that is, within a tolerance $r$, at the next point. Thus a low value of ApEn reflects a high degree of regularity. Importantly, the ApEn algorithm counts each sequence as matching itself, in the calculations to skip the occurrence of $\ln(0)$ a practice is carried over following the work of Eckmann and Ruelle \cite{Eckmann}. In practice, it is found that ApEn lacks two important expected properties. First, ApEn is heavily dependent on the record length and is uniformly lower than expected for short records. Second, it lacks relative consistency. That is, if ApEn of one data set is higher than that of another, it should, but does not, remain higher for all conditions tested \cite{Pincus9}. This shortcoming is particularly important, because ApEn has been repeatedly recommended as a relative measure for comparing data sets \cite{Pincus9,Pincus5}.

Following the approach of Grassberger and co-researchers [26-29], Richman and Moorman \cite{Richman} have developed sample entropy ($SampEn(m, r, N)$) which is precisely the negative natural logarithm of the conditional probability that two sequences similar for m points remain similar at the next point, where self-matches are excluded during the calculation of the probability. Thus a lower value of SampEn also indicates more self-similarity in the time series. In addition to eliminating self-matches, the SampEn algorithm is simpler than the ApEn algorithm, requiring approximately one-half as much time to calculate. SampEn is largely independent of record length and displays relative consistency under circumstances where ApEn does not. 

\begin{figure}[h!]
\begin{center}
\includegraphics[width= 0.73\textwidth]{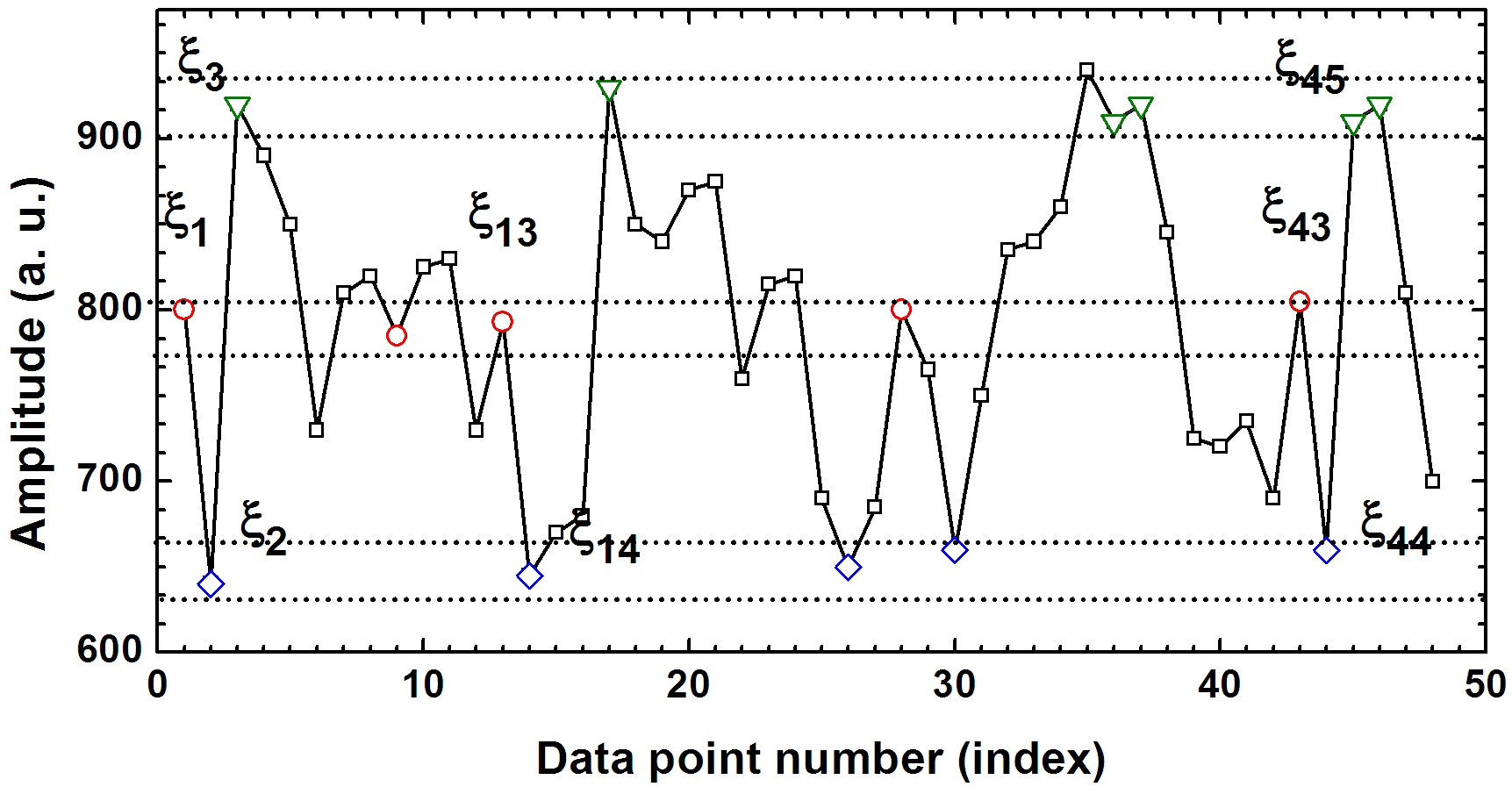}
\vspace{10pt}
\caption{A typical time series of 48 data points [$\xi_1,\xi_2,....,\xi_{48}$] is considered for the illustration of the SampEn calculation procedure for $m=2$ and for a real positive $r$ value. This simulated time series provides 47 and 46 two component and three component vector sequences respectively. The dotted horizontal lines around data points $\xi_1$, $\xi_2$, $\xi_3$ are $\xi_1\pm r$, $\xi_2\pm r$, $\xi_3\pm r$ lines respectively. The data points which match with first three data points($\xi_1$, $\xi_2$, $\xi_3$) are represented by symbols $\circ$, $\diamond$, and $\triangledown$ respectively.  For first two-component $\circ$-$\diamond$ vector[$\xi_1$, $\xi_2$] we find only two other matching $\circ$-$\diamond$ sequences [$\xi_{13}$, $\xi_{14}$] and [$\xi_{43}$, $\xi_{44}$]. This procedure is repeated for all the 47 two-component vectors and the matching counts for each two-component vectors are added up. Similarly for first three-component $\circ$-$\diamond$-$\triangledown$ vector[$\xi_1$, $\xi_2$, $\xi_3$] we find only one matching $\circ$-$\diamond$-$\triangledown$ sequence [$\xi_{43}$, $\xi_{44}$, $\xi_{45}$]. For all the 46 three-component vectors the matching procedure is repeated and the matching counts for each three-component vectors are added up. The natural logarithm of the ratio of total number of two-component matching and three-component matching provides the SampEn($m=2$, $r$) for that particular simulated time series(it is exactly the procedure suggested by Costa \emph{et~al.} in reference \cite{Costa2} for the calculation of SamEn of a given time series).}
\vspace{10pt}
\label{figg1}
\end{center}
\end{figure}

\textbf{\textit{Construction of SampEn}}:
If $ n_j^m(r)$ be the number of vectors $u(k)$ with in the distance $r$ of any particular vector $u(j)$($j \neq k$) the probability of that any $u(k)$ is within $r$ of $u(j)$ is $P^m_j(r)=n_j^m(r)/(N-m)$. $P^m(r)$ is defined by 
\begin{equation}  
 P^m(r)=1/(N-m)\sum^{N-m}_{j=1} P^m_j(r)
\end{equation} 
and the SampEn is defined as 
\begin{align}  
 SampEn(m,r,N) &=\ln \frac{P^m(r)}{P^{m+1}(r)}\nonumber\\
&=\ln \frac{\sum_{j=1}^{N-m}n_{j}^m(r)}{\sum_{j=1}^{N-m}n_{j}^{m+1}(r)}
\end{align} 
whereas,
\begin{align}  
ApEn(m,r,N) &=U^m(r)-U^{m+1}(r)\nonumber\\
&\approx \frac{1}{N-m}\sum _{j=1}^{N-m}\ln \frac{P_j^{\prime m}(r)}{P_j^{\prime (m+1)}(r)}\nonumber\\
& =\frac{1}{N-m}\sum _{j=1}^{N-m}\ln \frac{n_j^{\prime m}(r)}{n_j^{\prime (m+1)}(r)}
\end{align}
where $n_j^m$ differs from $n^{\prime m}_j$ to the extend that for SampEn self-matches are not counted $(j\ne k)$ and $1\le j \le N-m$. A typical example of the procedure for calculating SampEn ($m=2$ and $r$ is a arbitrarily chosen positive number) is illustrated in Fig.~\ref{figg1}.

Application of both the ApEn and SampEn algorithms assign higher entropy for certain pathologic time series data than free running healthy physiological data which is a bit confounding. Intuitively a pathologic time series represents less complex system and do not comply with numerical results. The reason behind this unphysical result is that all these algorithms are based on single scale and only the uncertainty associated to the next new point is reflected in the entropy. As already stated that the structural richness and the complex behavior of a time series is significantly tied with inherent multiple spatio-temporal scale ApEn and SampEn algorithms do not account the features related to multiple scales other than the original scale. Zhang \cite{Zhang} has proposed an approach to take into account the multi-scale information for large noise free data. Obviously physiological and physical signals are bounded and are not noise free. 

\begin{figure*}[ht!]
\begin{center}
\includegraphics[width= 0.90\textwidth]{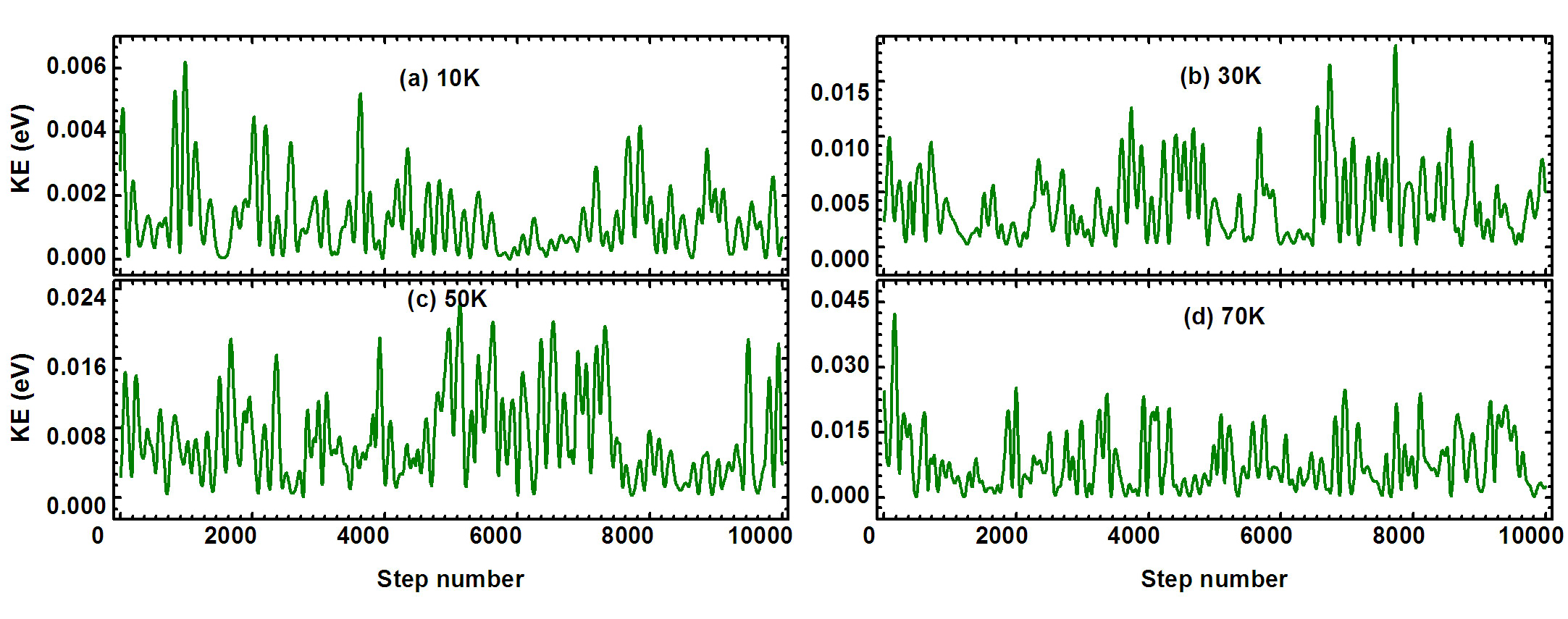}
\vspace{10pt}
\caption{Time series data of KE of single Ar atom at temperatures (a) 10K, (b) 30K, (c) 50K and (d) 70K. Each series contains $10^4$ data points.}
\vspace{10pt}
\label{figg2}
\end{center}
\end{figure*}

SampEn algorithm is free from these two limitations whereas Zhang's method takes into account the multi-scale effect. Thus, collective use of Pincus's, Richman's and Zhang's approach provides (treated as MSE analysis) the exact behavior of complex systems. Costa and co-workers used this approach, for the first time, to biological and physiological signals [2-10]. Afterwards, MSE analysis has been applied to metallurgical systems \cite{Asarkar1, Asarkar2} to calculate the complexity. In this paper we, for the first time, use this method to the time series data of KE  of individual atoms of solid Ar at various equilibrium temperatures. Fig.~\ref{figg2}(a)-\ref{figg2}(d) are the time series data of KE of an arbitrary atom at 10K, 30K, 50K and 70K respectively. 

\begin{figure}[h!]
\begin{center}
\includegraphics[width= 0.70\textwidth]{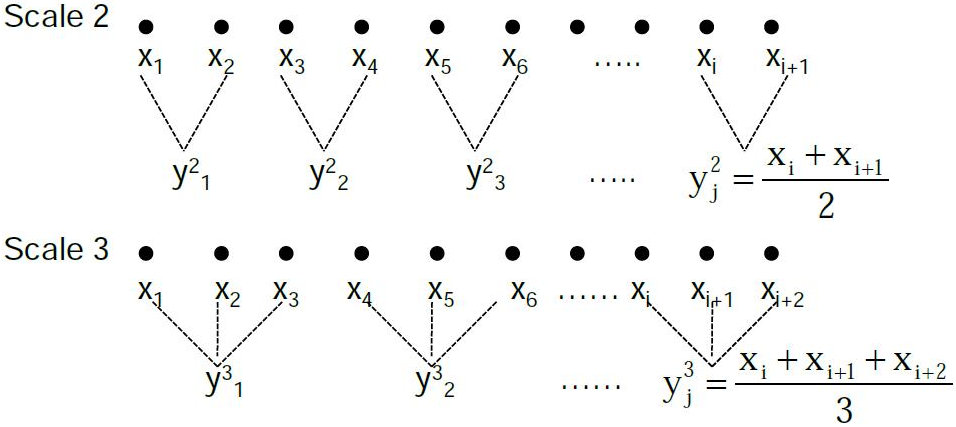}
\vspace{10pt}
\caption{Procedure to generate new coarse-grained time series of scale factor two and three from the original time series ($\{x_i\}=[x_1,x_2,...., x_N]$) which is also a coarse-grained time series with scale factor one.}
\vspace{10pt}
\label{figg3}
\end{center}
\end{figure}

\textbf{\textit{Construction of coarse-grained time series for different scale factors }}:
 Consider a finite time series data of $N$ points given by $\{x_i\}=[x_1,x_2,...., x_N]$. The algorithm for the coarse-grained time series corresponding to scale factor $\tau$ can be written as:
\begin{equation}  
  y^\tau_j = \frac{1}{\tau}\sum^{j\tau}_{i=(j-1)\tau+1} x_i 
\end{equation} 

where $y^\tau_j$ is the $j_{th}$ component of coarse-grained time series (\{$y^\tau_j$\}=[$y^\tau_1$,$y^\tau_2$,...,$y^\tau_{jmax}$]) with scale factor $\tau$ and $1\leq j\leq N/\tau$. Thus the coarse-grained series for scale factor $\tau=1$ is the original time series. Fig.~\ref{figg3} is a schematic representation of coarse-graining procedure for scale factor two and three.  Application of SampEn to each of these coarse-grained time series provides the MSE. 
\begin{figure}[h!]
\begin{center}
\includegraphics[width= 0.70\textwidth]{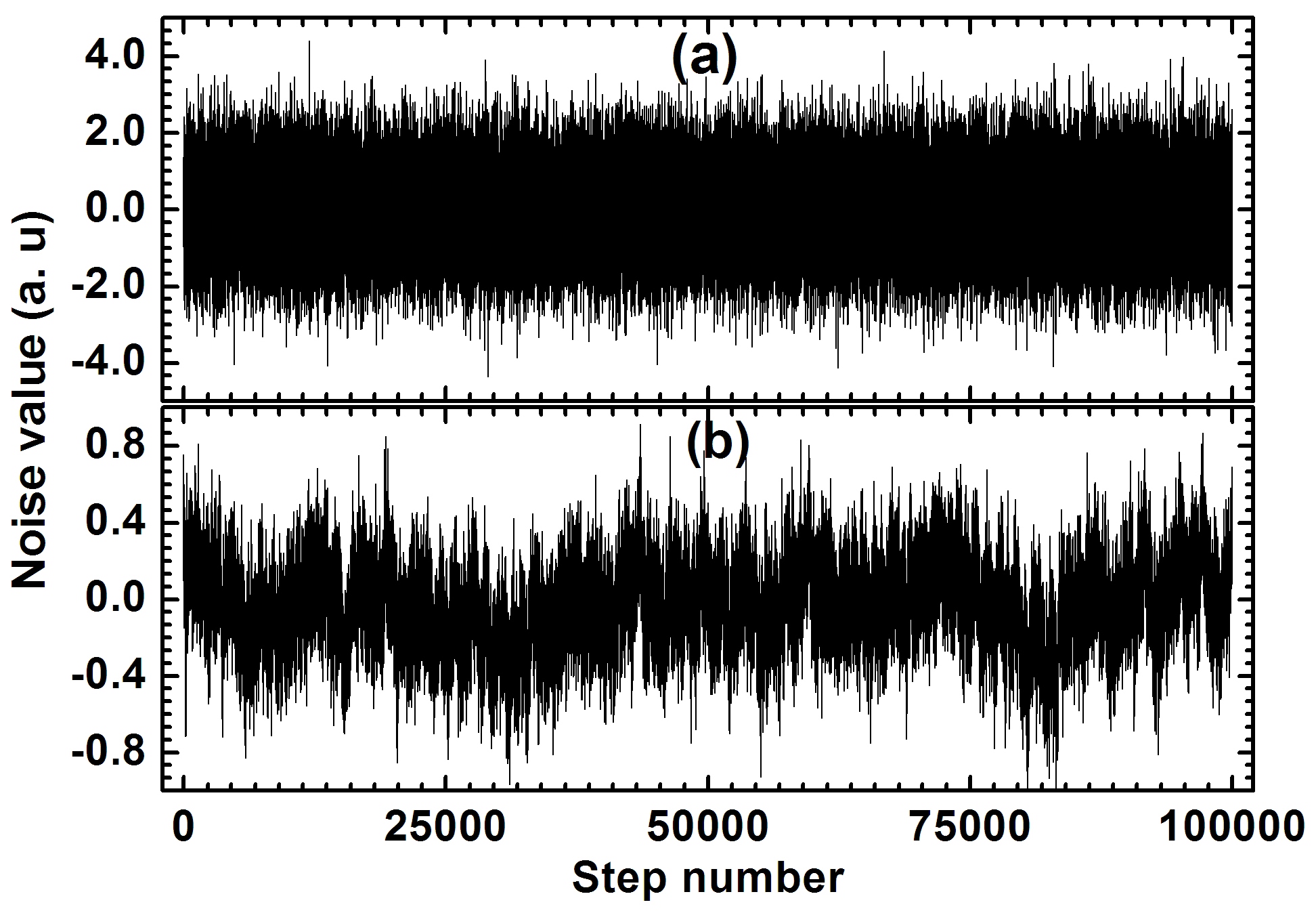}
\vspace{10pt}
\caption{Time series of (a) white noise and (b) 1/f noise with $10^5$ data points.}
\vspace{10pt}
\label{figg4}
\end{center}
\end{figure} 

\begin{figure}[h!]
\begin{center}
\includegraphics[width= 0.75\textwidth]{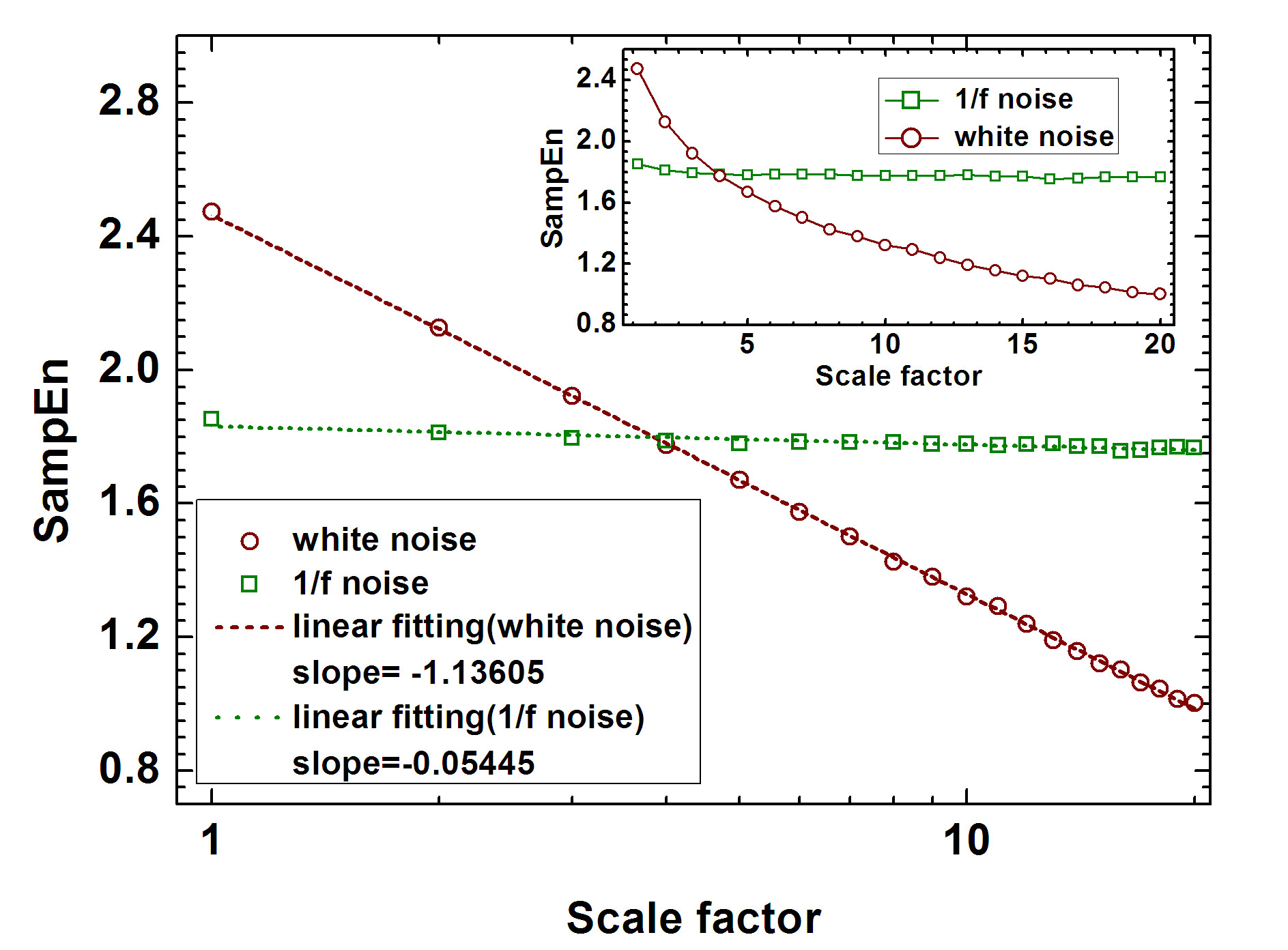}
\vspace{10pt}
\caption{MSE analysis($m=2$, $r=15\%$ of the standard deviation of the respective original time series) of white noise (circular) and 1/f noise (square) with $10^5$ data points showing the logarithmic relation between SampEn and scale factor and the inset is exactly the result showed by Costa \emph{et~al.} \cite{Costa1, Costa2}}
\vspace{10pt}
\label{figg5}
\end{center}
\end{figure}

\begin{figure}[h!]
\begin{center}
\includegraphics[width= 0.70\textwidth]{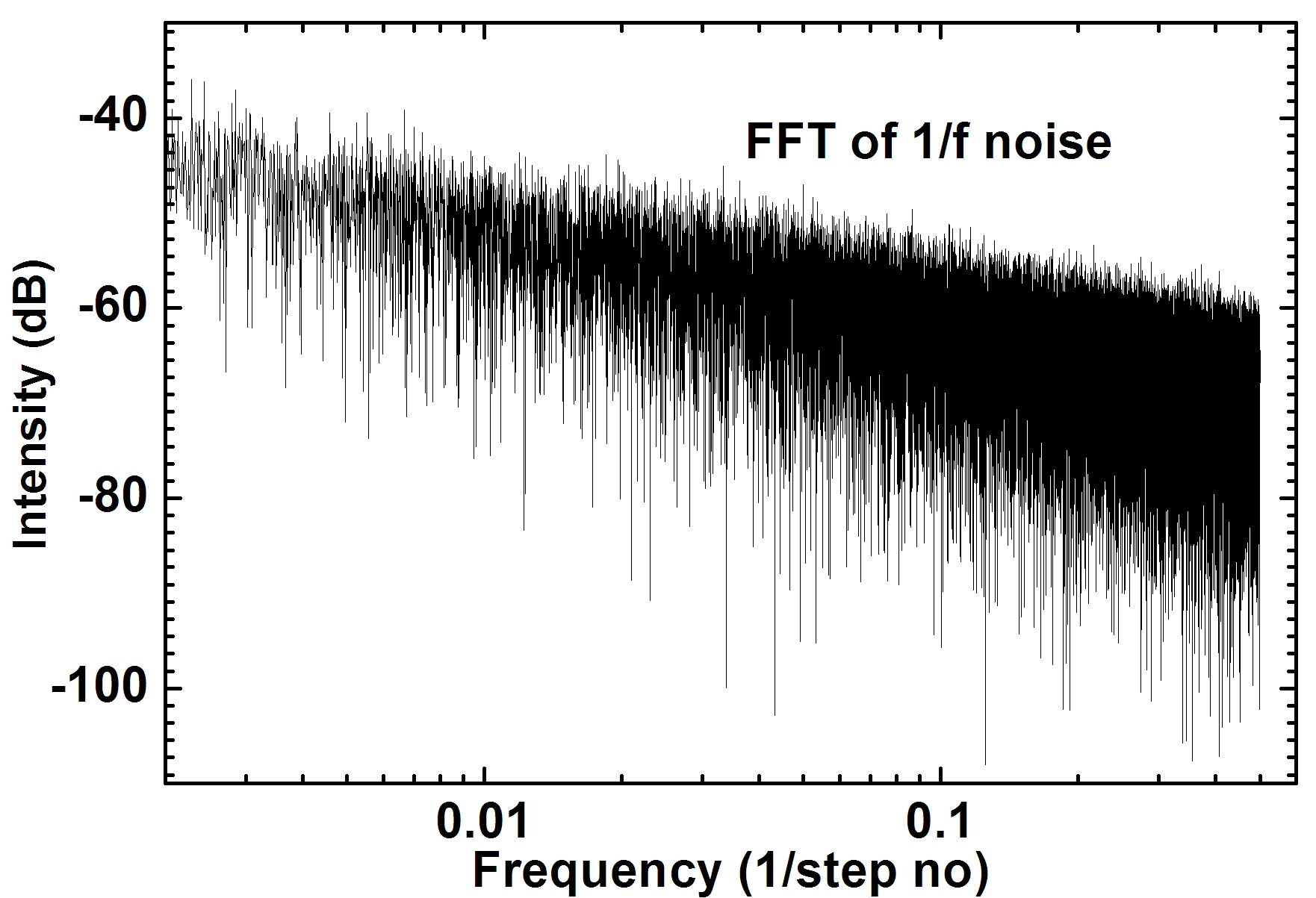}
\vspace{10pt}
\caption{Frequency spectrum of 1/f noise.}
\vspace{10pt}
\label{figg6}
\end{center}
\end{figure}

This method is applied to white noise (Fig.~\ref{figg4}(a)) and $1/f$ (Fig.~\ref{figg4}(b)) noise (with $10^5$ data points) to reproduce the result (inset of Fig.~\ref{figg5}) which was reported by Costa and co-researchers \cite{Costa1, Costa2}. Fig.~\ref{figg5} 
shows a congruity between numerical results and our intuition about regularity and complexity of these two types of noise. White noise contains all the frequencies with equal probability. Hence no constraint will be there on the similarity of the frequency of data points of white noise. This regularity is reflected in the lower values of multi-scale entropy. On the contrary for 1/f noise lower frequencies are most likely compared to the higher frequencies (Fig.~\ref{figg6}). This particular restriction makes the 1/f noise to lose the regularity in the data points. The higher values of SampEn justifies this idea. The variation of SampEn (Fig.~\ref{figg5}) against \textit{log} scale is linear for white noise with negative slope -1.13605. Whereas, for 1/f noise the variation is not exactly linear but slightly tilted for smaller values of scale factors(scale factor 1-3). With a rough estimation, the linear fitting of the variation of SampEn of 1/f noise against \textit{log} scale gives a near zero slope of -0.05445.\\\\
{\bf 4. Simulation procedure}\\
The required time series data, for the measurement of complexity of crystalline systems, are generated with the help of molecular dynamics simulation technique. Molecular dynamics provides the time series data of KE of the individual constituent atoms in the time interval of the order of femtosecond(fs). A cubic system of solid Ar of dimension of 30 unit cells(uc) in each of the three directions is taken as simulation cell. The crystalline solid Ar has face centered cubic structure and the simulation cell contains $108000$ atoms. Periodic boundary condition is imposed in all the three directions of the cell to avoid any surface effect. At every 2fs time interval the Newtonian equations of motion of each constituent atom of the system are solved with the help of widely used Lennard-Jones(12-6) potential \cite{Lennard-Jones}.
\begin{figure}[h!]
\begin{center}
\includegraphics[width= 0.70\textwidth]{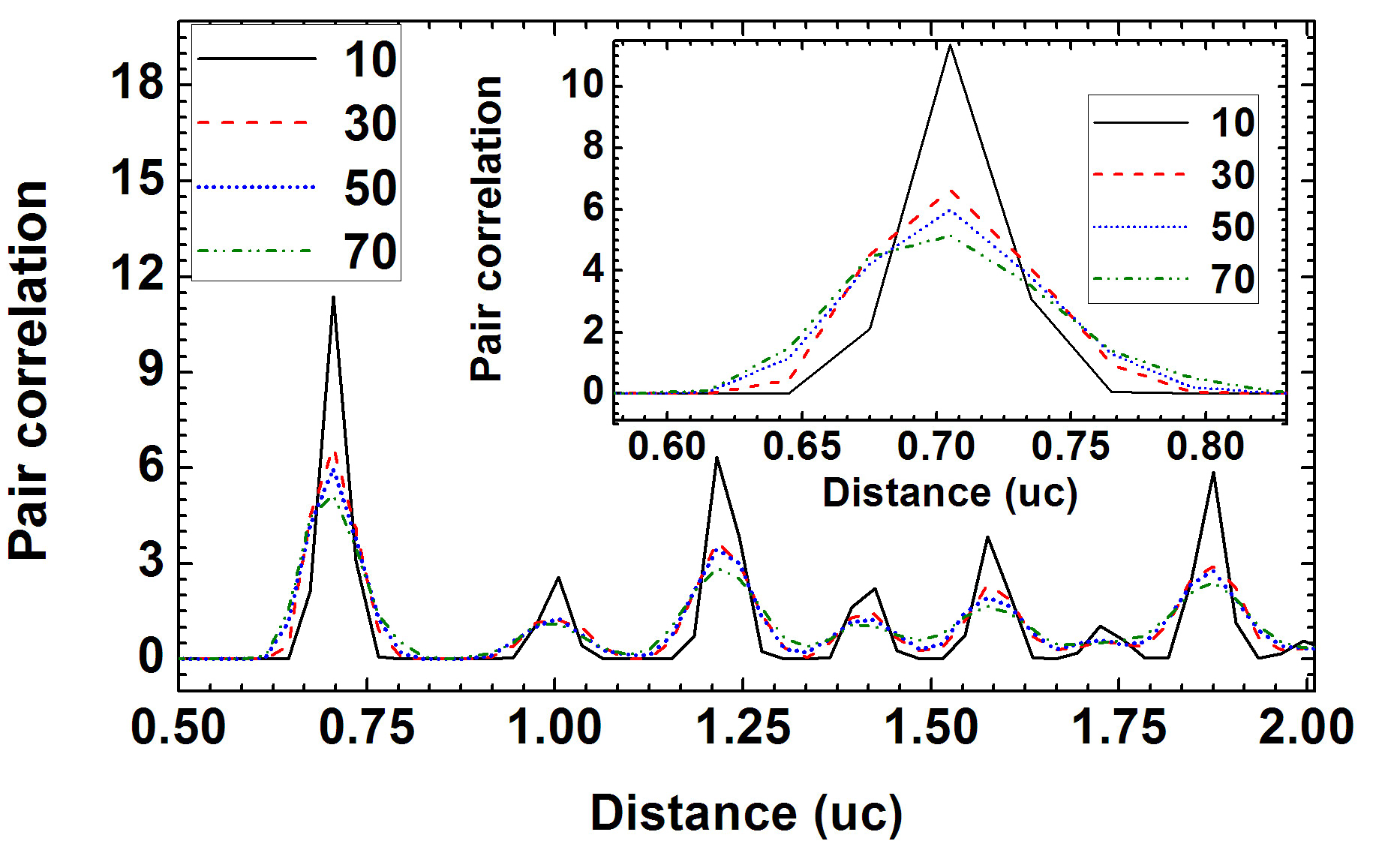}
\vspace{10pt}
\caption{Variation of pair correlation function of solid Ar at different equilibrium temperatures. The figure in the inset represents the enlarged view of the first peak.}
\vspace{10pt}
\label{figg7}
\end{center}
\end{figure} 

 Initially, the velocity components of all the constituent atoms of the simulation cell are defined by random numbers such that the initial average KE of the system becomes double of the value what we are looking for in equilibrium and the average potential energy(PE) is zero. Thereafter, using constant energy and volume ensemble (NVE) the crystalline solid Ar system is left of its own for long time (50000 steps, 100ps) to reach the equilibrium distribution. Under equilibrium, the final average KE is equal to the average PE and because of this transfer of KE into PE, the final average KE value reaches the desired value (half of the initial average KE). After the system reaches the equilibrium, the simulation run is continued for another 20ps and atomistic simulation data are recorded at each time step of that 20ps (10000 steps) interval to study the complex dynamics of the equilibrium crystalline system. For different equilibrium temperatures (10, 30, 50 and 70K) of crystalline solid Ar system this procedure is repeated. \\\\
{\bf 5. Results and discussions}\\
In Fig.~\ref{figg7} the pair correlation function (sometimes called radial distribution function) of the Ar atoms is portrayed for different equilibrium temperatures. Pair correlation function describes how the normalized particle density varies as a function of distance from a reference particle. The atom residing at the center of the simulation cell is considered as the reference particle. For closer distance it exhibits peaks at first, second, third nearest neighbor etc. and gradually flattens to unity for larger distance. 
\begin{figure}[h!]
\begin{center}
\includegraphics[width= 0.70\textwidth]{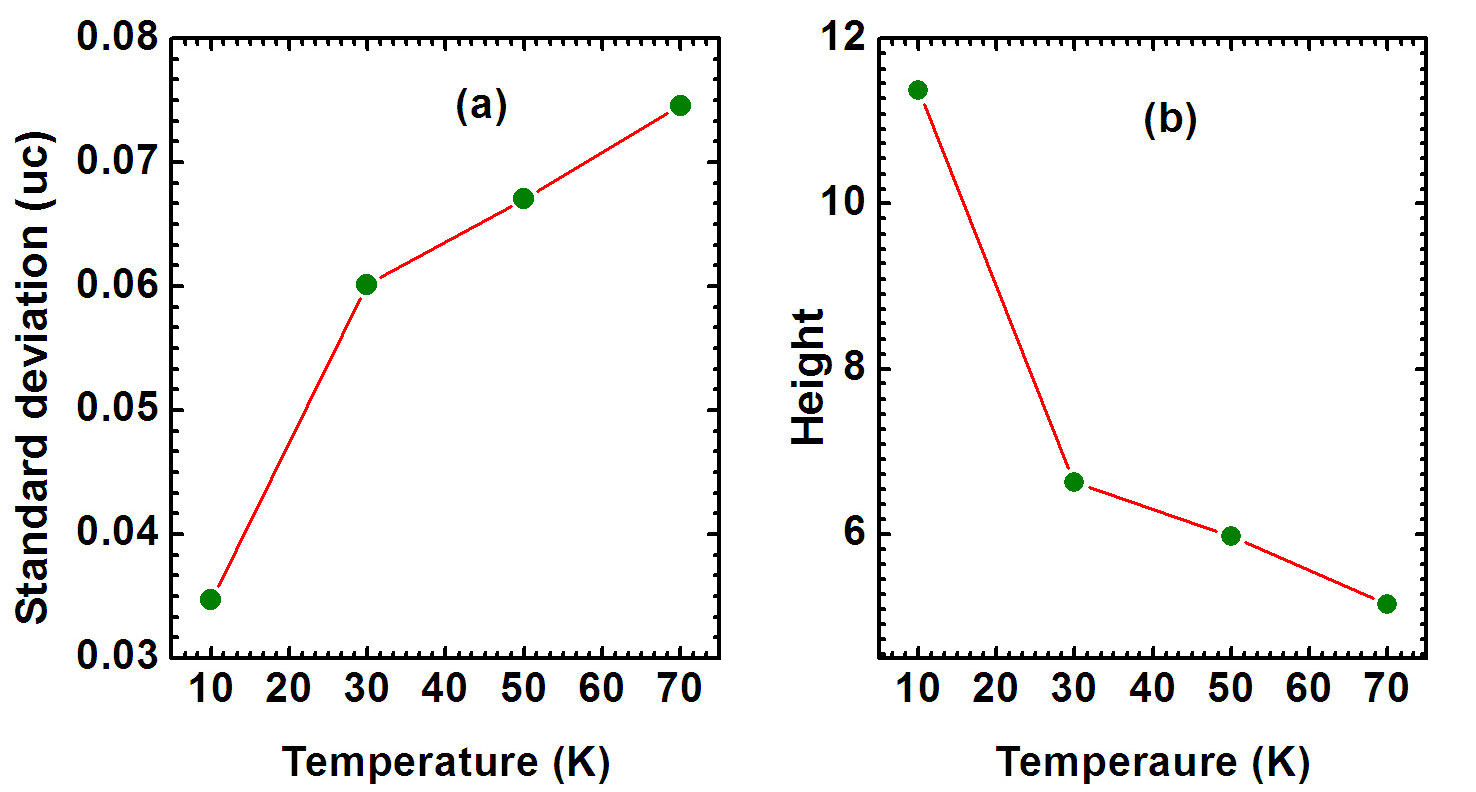}
\vspace{10pt}
\caption{Temperature variation of (a) standard deviation and (b) height of the first peak.}
\vspace{10pt}
\label{figg8}
\end{center}
\end{figure}
An increment in temperature enhances the KE of individual atoms and hence the amplitude of vibration about their respective mean positions. Gradual flattening of the first peak at the cost of peak height (in the inset of Fig.~\ref{figg7}) consolidate the concept of larger KE and larger vibration of atoms for increasing temperatures. Thus, rise of temperature introduces more randomness and irregularity to the dynamics of system particles. Temperature dependence of the standard deviation and the height of the first peak of pair correlation function are shown in Fig.~\ref{figg8}(a) and Fig.~\ref{figg8}(b) respectively.

\begin{figure}[h!]
\begin{center}
\includegraphics[width= 0.70\textwidth]{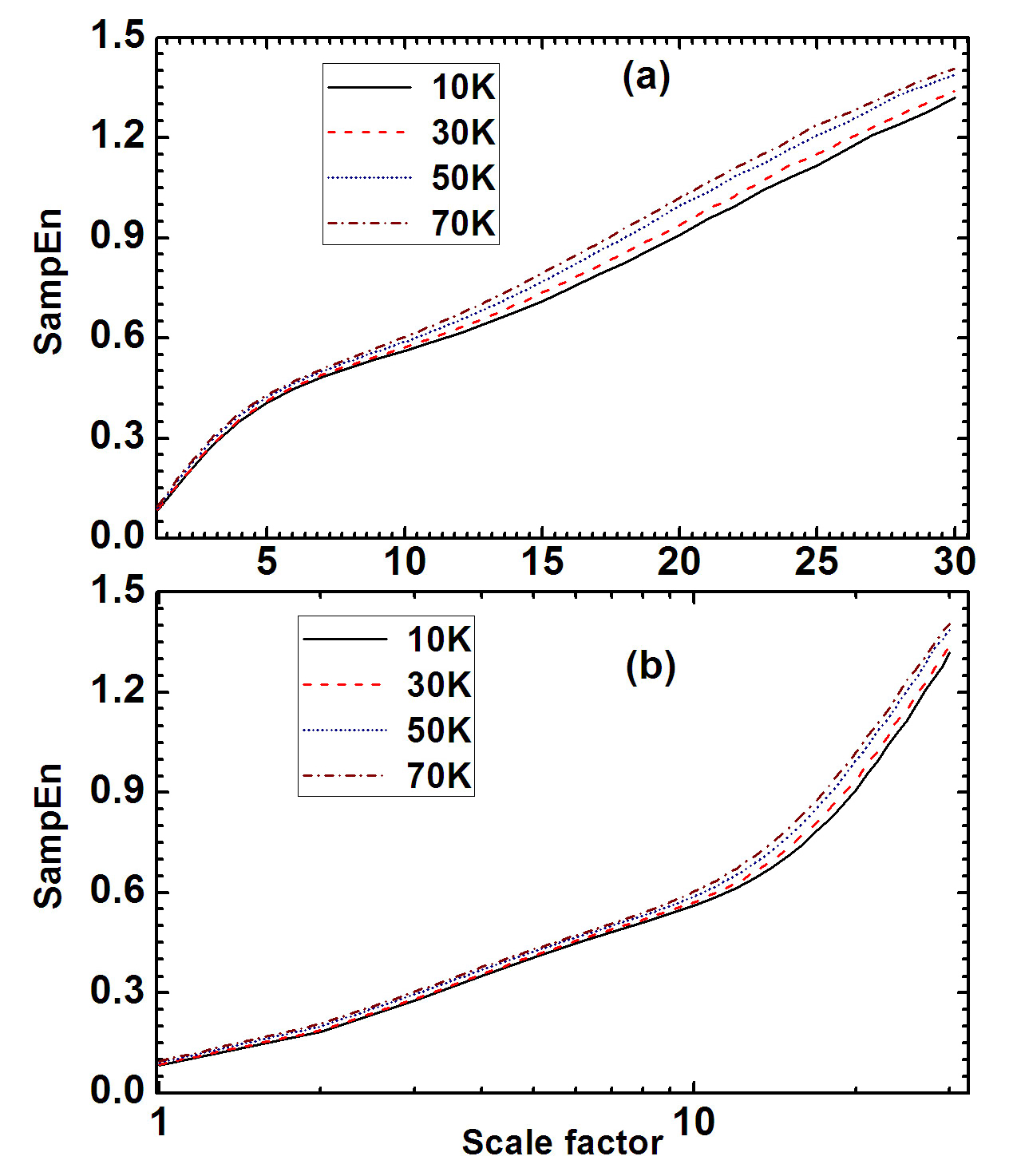}
\vspace{10pt}
\caption{Calculated SampEn for different equilibrium temperatures of solid Ar in (a) \textit{linear} scale and in (b) \textit{log} scale. }
\vspace{10pt}
\label{figg9}
\end{center}
\end{figure}

\begin{figure}[h!]
\begin{center}
\includegraphics[width= 0.55\textwidth]{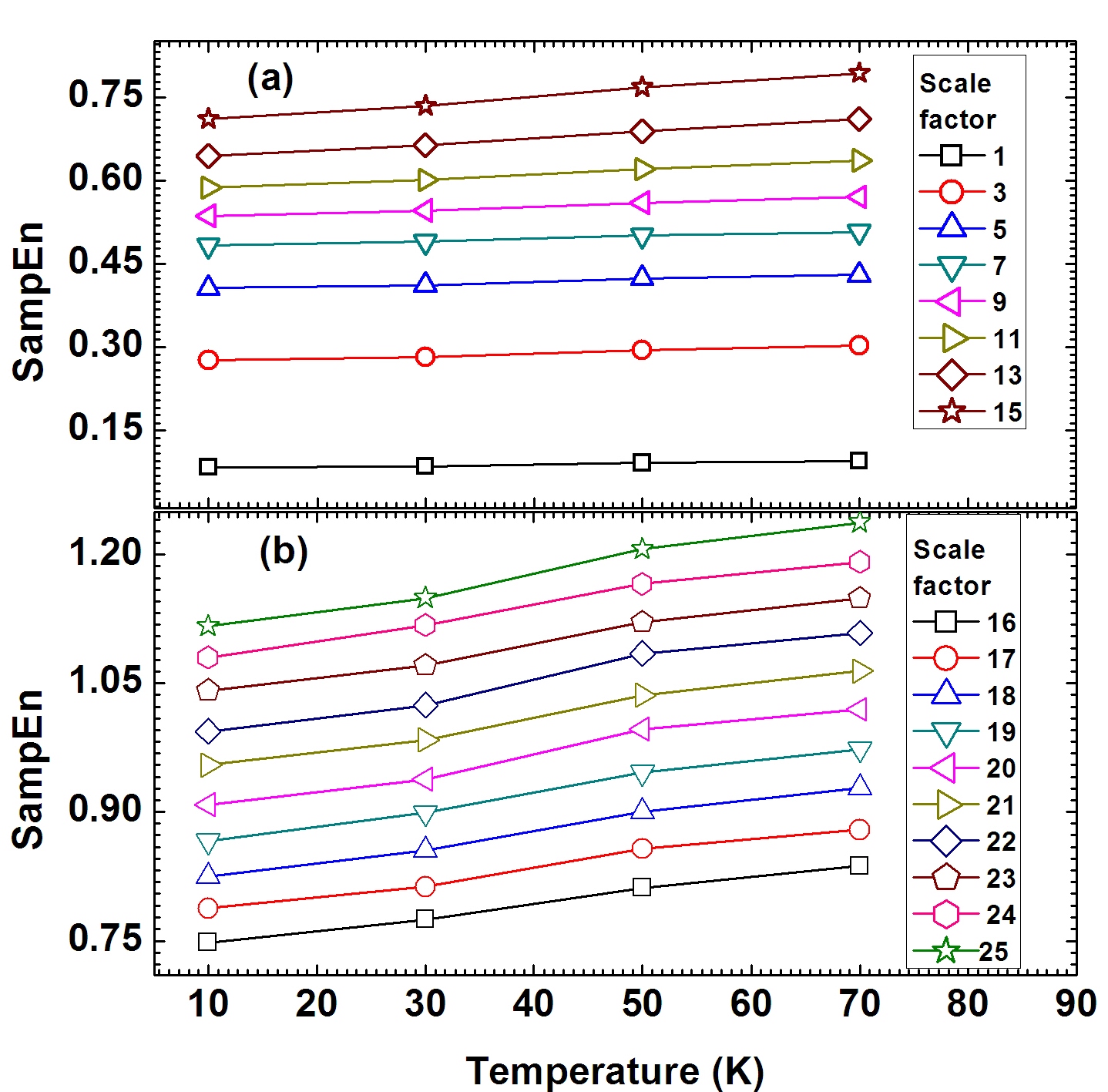}
\vspace{10pt}
\caption{Dependence of SampEn on temperature with (a) every alternate value of scale factors ranging from 1 to 15 and (b) for each value of scale factors ranging from 16 to 25.}
\vspace{10pt}
\label{figg10}
\end{center}
\end{figure}

 Few atoms (around 70) in the simulation cell are identified with in a sphere of radius 1.6uc whose center is chosen to be almost at the middle of the cell, far away from the surface. The time series data ($10^4$ data points) of KE of the identified atoms are used for the analysis. Throughout the analysis the values of SampEn are estimated with $m=2$ and the $r$ value equals to $15\%$ of the standard deviation of original time series data. Subsequently the average of SampEn over the identified atoms is considered for the quantification of the complex behavior of the particle dynamics. The dependence of average SampEn on scale factor is shown in Fig.~\ref{figg9} 
and the Fig.~\ref{figg10} provides an idea about the temperature variation of complex nature of the system for different scale factors ranging from 1-25. It is observed that for a particular scale factor SampEn increases linearly with temperature.

The pair correlation function (Fig.~\ref{figg7}) and the calculated SampEn (Fig.~\ref{figg9}) corroborate each other. Pair correlation function reflects the idea what our intuition tells about the complex dynamics of the system. Whereas, MSE is the method to quantify the irregularity of uncorrelated noise of a complex system. Flattening of pair correlation peak (inset of Fig.~\ref{figg7}) with the enhancement of temperature and the linear dependence of SampEn on temperature (Fig.~\ref{figg10}) both are the signature of analogous fact that temperature makes the system dynamics more uncorrelated, noisy and irregular.

\begin{figure}[h]
\begin{center}
\includegraphics[width= 0.70\textwidth]{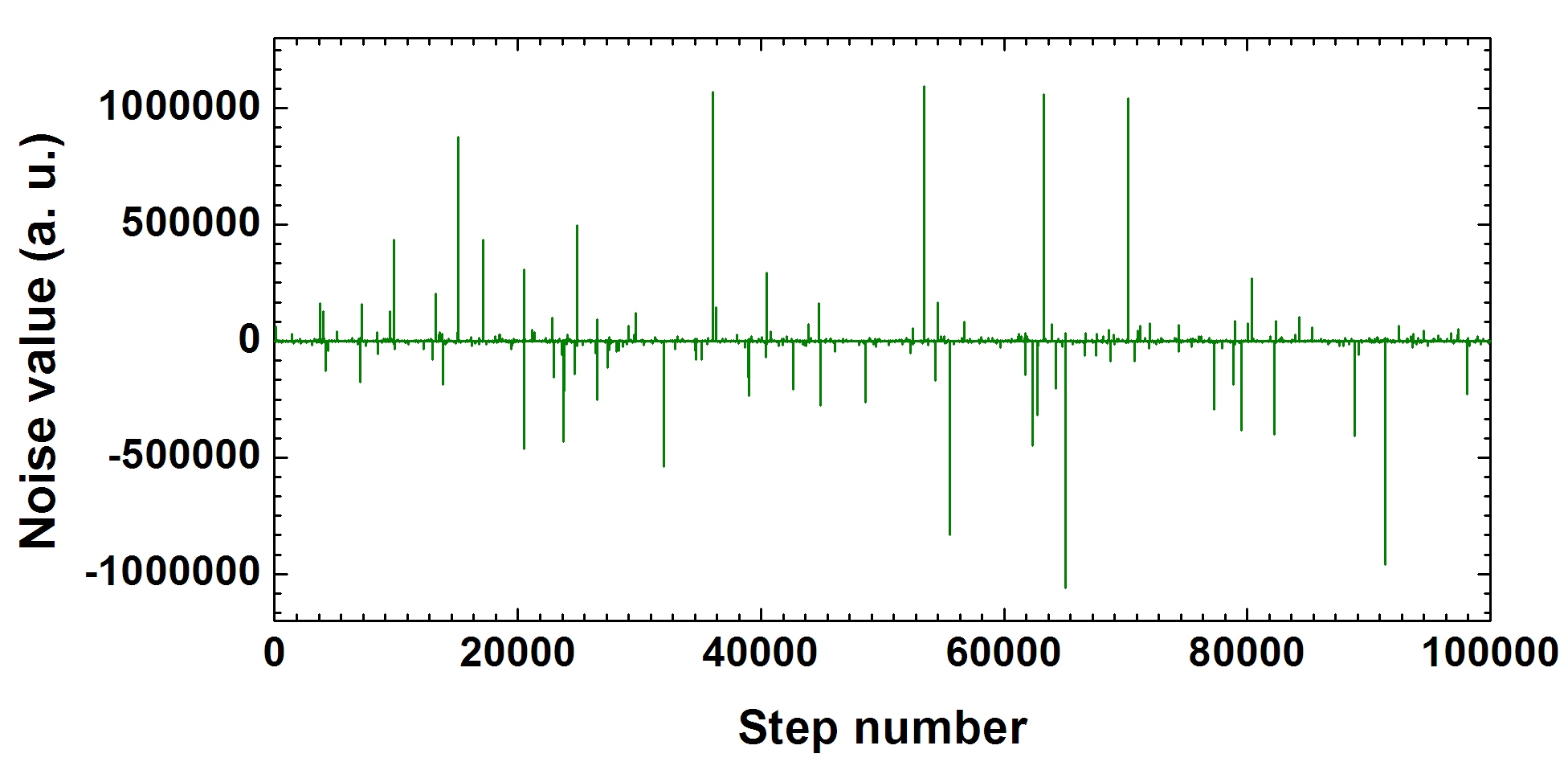}
\vspace{10pt}
\caption{A typical Levy noise with $10^5$ data points. Hurst exponent, $\alpha=0.7$}
\vspace{10pt}
\label{figg11}
\end{center}
\end{figure}

\begin{figure}[h!]
\begin{center}
\includegraphics[width= 0.65\textwidth]{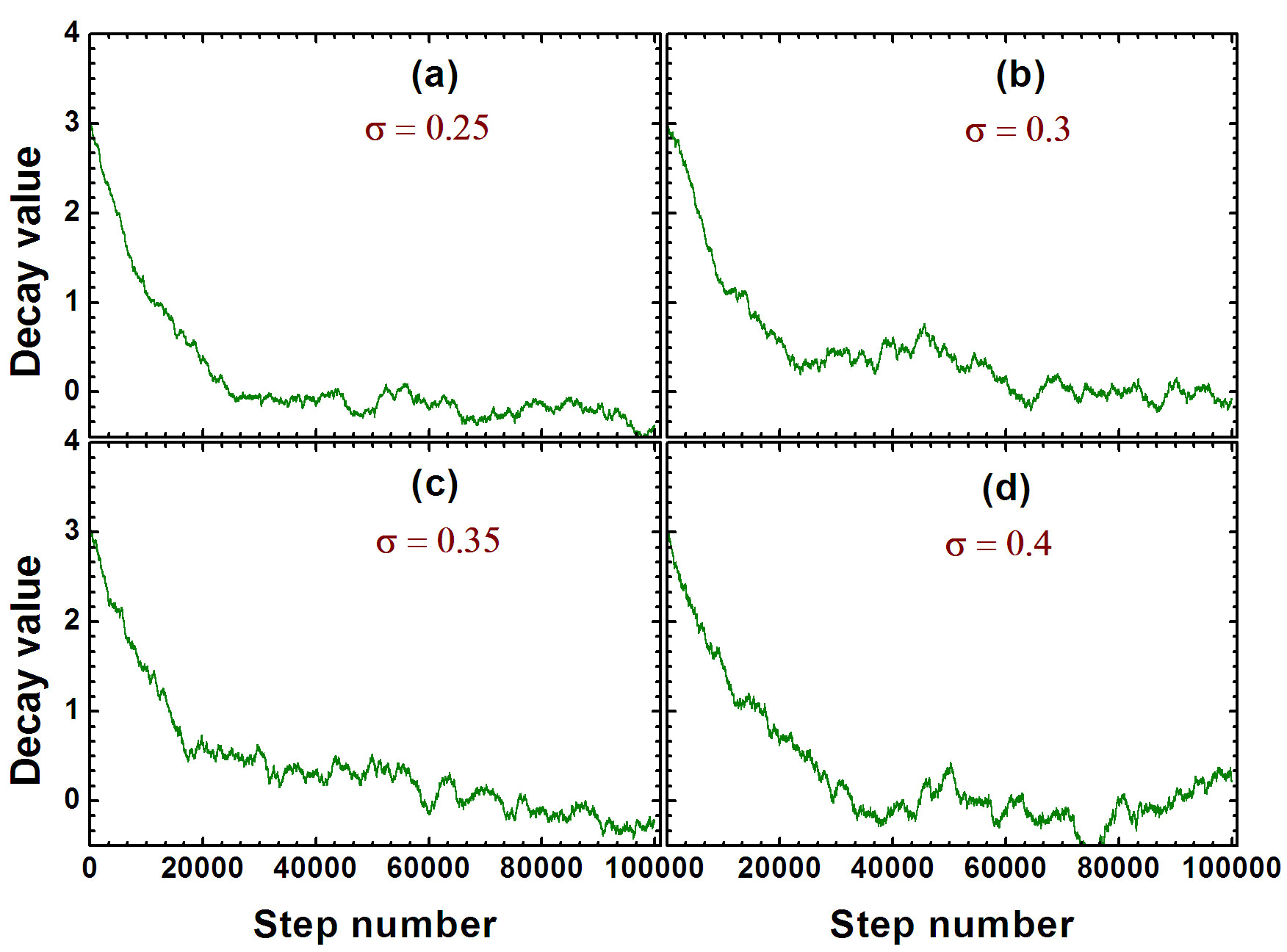}
\vspace{10pt}
\caption{Exponential decay series with different amplitude of the noise component. (a) $\sigma = 0.25$, (b) $\sigma = 0.3$, (c) $\sigma = 0.35$ and (d) $\sigma = 0.4$. Each series starts at initial value 3.}
\vspace{10pt}
\label{figg12}
\end{center}
\end{figure}

 Another important outcome of the current MSE analysis is the peculiar random nature of time series data of KE. Continuous coarse-graining procedure can not eliminate the uncorrelated random component of the time series which is originally uncorrelated and irregular. Rather the procedure progressively makes the correlation of the time series even worse. The inherent multiple spatio-temporal scale dependence is exposed for higher scale factors. Undoubtedly, only the asymptotic value of SampEn (scale factor one) does not make any sense about the complexity and the introduction of multi-scale factor in the analysis of complexity affirms its utility.\\\\
{\bf 6. Conclusions}\\
Recently Barat \emph{et~al.} \cite{Pbarat} have shown that in crystalline solid systems the time series data of the KE of individual atoms exhibits Levy walk property. On the other hand using Ford-Kac model \cite{Ford} and Caldeira-Leggett model \cite{Caldeira1, Caldeira2}, Hasegawa \cite{Hasegawa} has shown that the individual particles of a classical small system coupled to finite bath follow Langevin equation. We have synthetically generated the Levy noise and Langevin solution and an exercise has been carried out on the complex behavior of these noisy signals.  The Levy noise is the time series of $10^5$ data points having Hurst exponent 0.7 (Fig.~\ref{figg11}). The time series of Langevin solution represents an exponential decay with noise given by the equation: 

\[\,dx(t) =-x(t) \,dt+\sigma \,dB(t)\]

In discrete form

\[x(i+1) =x(i)-x(i)\Delta t+\sigma \sqrt{\Delta t}\times \mbox{(a~random~number)}\] 
where, the $\sigma$ value introduce the noise component to the decay series and $\Delta t$ is the dimensionless time interval. The decay constant being dimensionless unity. Eight exponential decay series with varying $\sigma$ values (0.05, 0.1, 0.15, 0.2, 0.25, 0.3, 0.35, 0.4) are generated (Fig.~\ref{figg12}). At time $t=0$ the decay starts with initial value 3 and data points are stored at every $\Delta t=8\times 10^{-5}$ time interval.        
\begin{figure}[h!]
\begin{center}
\includegraphics[width= 0.70\textwidth]{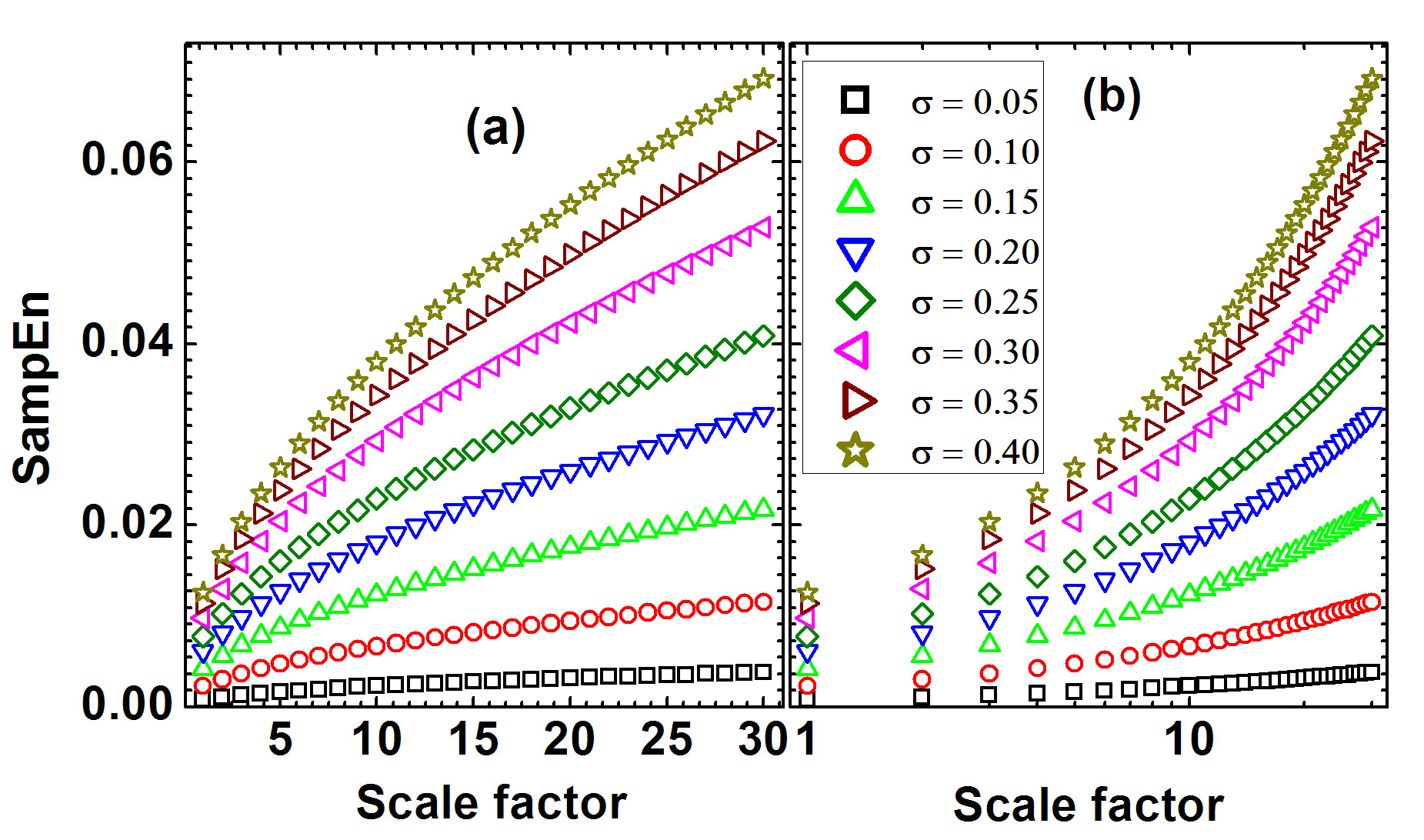}
\vspace{10pt}
\caption{MSE analysis of exponential decay series with noise(Langevin solution) in (a) \textit{linear} scale and in (b) \textit{log} scale.}
\vspace{10pt}
\label{figg13}
\end{center}
\end{figure}
\begin{figure}[h!]
\begin{center}
\includegraphics[width= 0.70\textwidth]{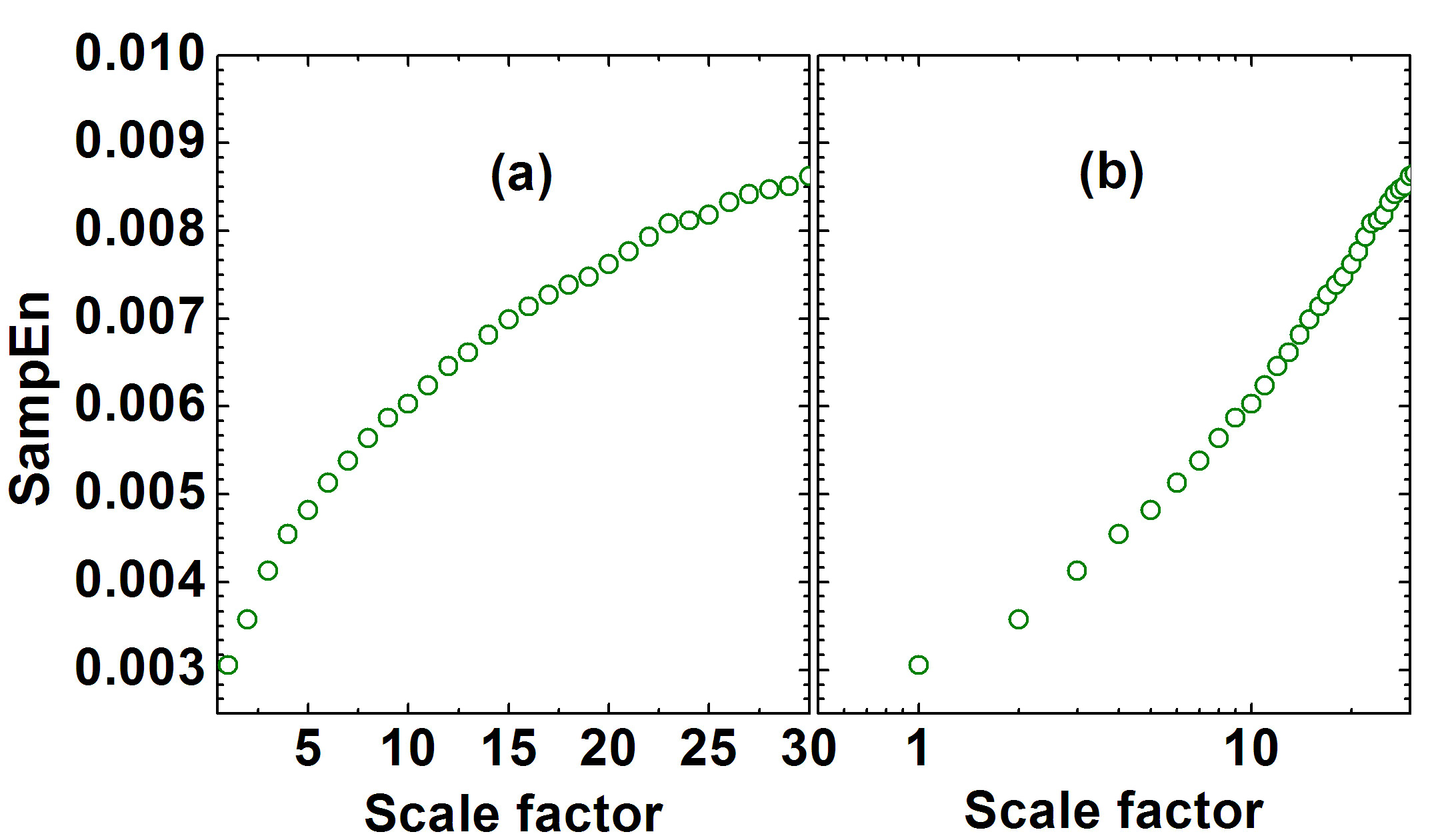}
\vspace{10pt}
\caption{MSE analysis of Levy noise in (a) \textit{linear} scale and in (b) \textit{log} scale.}
\vspace{10pt}
\label{figg14}
\end{center}
\end{figure} 

For MSE analysis of Langevin solutions only the flat tail parts of decay series (containing final $5\times10^4$ data points) are considered. The quantified SampEn values for Langevin solutions are illustrated in Fig.~\ref{figg13}. Similar to the temperature effect (enhancement of the amplitude of atomic vibrations) in solid Ar crystal the SampEn of decay series with larger $\sigma$ values quantifies higher degree of uncorrelated randomness. The scale dependence of SampEn for both Levy noise (Fig.~\ref{figg14}) and Langevin solutions exhibit similar sort of tendency as that of the time series data of KE of individual atoms in solid Ar system. Unlike the white noise and 1/f noise, in case of time series data associated to molecular dynamics generated of KE of Ar atoms, Levy noise, and Langvin solution the SampEn is not related to scale factor by logarithmic function. With scale factor, in all these three cases, SampEn increases with decreasing slope. Thus, the complexity analysis of time series data of KE strengthen the remarks of references \cite{Pbarat} and \cite{Hasegawa}.

\end{spacing}


\end{document}